\RequirePackage{ifpdf}
\ifpdf 
\documentclass[pdftex]{sigma}
\else
\documentclass{sigma}
\fi

\numberwithin{equation}{section}

\begin{document}

\allowdisplaybreaks

\renewcommand{\PaperNumber}{009}

\FirstPageHeading

\renewcommand{\thefootnote}{$\star$}

\ShortArticleName{Asymmetric Twin Representation: the Transfer Matrix
Symmetry}

\ArticleName{Asymmetric Twin Representation:\\ the Transfer Matrix
Symmetry\footnote{This paper is a contribution 
to the Vadim Kuznetsov Memorial Issue ``Integrable Systems and Related Topics''.
The full collection is available at 
\href{http://www.emis.de/journals/SIGMA/kuznetsov.html}{http://www.emis.de/journals/SIGMA/kuznetsov.html}}}

\Author{Anastasia DOIKOU}

\AuthorNameForHeading{A. Doikou}

\Address{INFN Section of Bologna, Physics Department, University of Bologna,\\
Via Irnerio 46, 40126 Bologna, Italy}
\Email{\href{mailto:doikou@bo.infn.it}{doikou@bo.infn.it}} 

\ArticleDates{Received August 02, 2006, in f\/inal form
December 26, 2006; Published online January 09, 2007}

\Abstract{The symmetry of the Hamiltonian describing the asymmetric
twin model was partially studied in earlier works, and our aim here
is to generalize these results for the open transfer matrix. In this
spirit we f\/irst prove, that the so called boundary quantum algebra
provides a symmetry for any generic~-- independent of the choice of
model~-- open transfer matrix with a trivial left boundary. In
addition it is shown that the boundary quantum algebra is the
centralizer of the $B$ type Hecke algebra. We then focus on the
asymmetric twin representation of the boundary Temperley--Lieb
algebra. More precisely, by exploiting exchange relations dictated
by the ref\/lection equation we show that the transfer matrix with
trivial boundary conditions enjoys the recognized ${\cal
U}_{q}(sl_2) \otimes {\cal U}_{{\mathrm i}}(sl_2)$ symmetry. When 
a~non-diagonal boundary is implemented the symmetry as expected is
reduced, however again certain familiar boundary non-local charges
turn out to commute with the corresponding transfer matrix.}


\Classification{81R50; 17B37}

\section{Introduction}

As is well known Yang--Baxter \cite{baxter, korepin} and ref\/lection
equations \cite{cherednik, sklyanin} provide sets of algebraic
constraints exactly def\/ining integrable models with periodic and
open boundary conditions respectively (see e.g.~\cite{sklyanin,
takt}). On the other hand it was argued \cite{jimbo, drinf} that the
algebraic structures underlying Yang--Baxter equation may be seen as
deformations of the usual Lie algebras, called quantum algebras. In
a similar fashion the ref\/lection algebras arise naturally in the
frame of the ref\/lection equation \cite{sklyanin}, and they have been
the subject of ongoing interest (see e.g.~\cite{mene,mr,dema,doikoun,doikou2}). The f\/irst step to comprehend such
algebras is to systematically classify the solutions of the
Yang--Baxter and ref\/lection equations. This is a major problem, the
study of which will signif\/icantly contribute into the deep
understanding of the mathematical mecha\-nisms that rule integrable
models with boundaries. A consistent scheme to derive solutions of
the ref\/lection equation is to exploit the structural similarity
between the Yang--Baxter and ref\/lection equations with the cylinder
braid group \cite{lema, lema1, ourpaper}. Then, using elements of
representation theory one may construct physical systems based on
the aforementioned solutions.

One of the main challenges within this context is to determine exact
symmetries associa\-ted to the physical systems under consideration.
In the present article we use already known representations of the
boundary Temberley--Lieb (blob) algebra \cite{lema, lema1, ourpaper}
as solutions of the ref\/lection equation, and we derive explicitly
conserved non-local charges belonging to the underlying ref\/lection
algebra, for various choices of boundary conditions. In particular,
the symmetry of the transfer matrix of the so called asymmetric twin
model is investigated. The main motivation for studying this model
is the fact that it is a novel system altogether, and as such it
of\/fers a rather unconventional perspective on boundary ef\/fects.
Historically the asymmetric twin representation was introduced in
\cite{MartinWoodcock01pre}, whereas the corresponding physical model
was constructed and studied in \cite{ourpaper, our2}. It was shown
in \cite{our2} that certain quantities were the centralizers of the
asymmetric twin representation of the boundary Temperley--Lieb
algebra. The corresponding Hamiltonian was then expressed in terms
of the generators of the blob algebra, hence it turned out to
commute with the aforementioned centralizers. In addition the form
of the spectrum and the Bethe ansatz equations for the asymmetric
twin transfer matrix were determined via the analytical Bethe ansatz
formulation exploiting also the spectral equivalence between the
open XXZ and asymmetric twin Hamiltonians. In the present work with
the help of exchange relations satisf\/ied by solutions of the
ref\/lection equation we are able to identify quantities commuting
with the transfer matrix of the model. The transfer matrix as known
gives rise to a~whole hierarchy of charges in involution rendering
the system integrable, with the Hamiltonian being the f\/irst one of
the hierarchy.

The organization of this article is as follows: In the next section
we introduce quotients of the cylinder braid group, i.e.\ the $B$
type Hecke algebra and the boundary Temperley--Lieb algebra. We also
review representations of the boundary Temperley--Lieb algebra, that
is the XXZ and asymmetric twin representations. In Section~3 the
transfer matrix of an open spin chain is reviewed. In Section~4 the
underlying ref\/lection algebra is introduced. It is also shown that
the boundary quantum algebra~-- emerging from the ref\/lection algebra
with no spectral parameter~-- provides in fact a symmetry for any
generic open transfer matrix with a trivial left boundary, and it is
the centralizer of the $B$ type Hecke algebra as well. The
implementation of a non trivial left boundary breaks the symmetry
down to a consistent subset. Note that these f\/indings are generic
that is they are independent of the choice of model. In the next
section basic def\/initions regarding the quantum algebra ${\cal
U}_q(\widehat{sl_{2}})$ are reviewed, and generalized intertwining
relations are introduced. The symmetry for both the open XXZ and
asymmetric twin models is also studied with the help of the
aforementioned intertwining relations. The XXZ chain is considered
mostly as a warm up exercise, however the main aim in this section
is the derivation of familiar conserved quantities for the open
asymmetric twin model for diagonal and non-diagonal right boundary.
It will become clear later that the intricate asymptotic behavior of
the asymmetric twin model \cite{our2} obliges  us to invoke exchange
relations involving solutions of the ref\/lection equation in order to
extract recognized conserved quantities. Finally in the last section
we discuss the main results of the present work, and we also give
some ideas for possible future directions.

\renewcommand{\thefootnote}{\arabic{footnote}}
\setcounter{footnote}{0}

\section[The $B$ type Hecke and blob algebras]{The $\boldsymbol{B}$ type Hecke and blob algebras}

It will be useful to f\/irst introduce a quotient of the cylinder
braid group algebra called $B$ type Hecke algebra ${\cal
B}_N(q,\delta_{e},\kappa)$, def\/ined by generators ${\mathbb U}_{1},
{\mathbb U}_{2}, \ldots, {\mathbb U}_{N-1}$ and $e$, and relations:
\begin{gather} {\mathbb U}_{i} {\mathbb U}_{i} = \delta {\mathbb U}_{i}, \nonumber\\
{\mathbb U}_{i} {\mathbb U}_{i+1} {\mathbb U}_{i} - {\mathbb
U}_{i}= {\mathbb U}_{i+1} {\mathbb U}_{i} {\mathbb U}_{i+1}
-{\mathbb U}_{i+1}, \nonumber\\ \big [ {\mathbb U}_{i}, {\mathbb U}_{j}
\big ] = 0, \qquad |i-j| \neq 1 \label{TL} \end{gather} (so far we
have the ordinary Hecke algebra ${\cal H}_{N}(q)$) with
$-\delta=q+q^{-1}$, $q=e^{{\mathrm i}\mu}$
\begin{gather}
e e = \delta_{e} e, \nonumber\\
{\mathbb U}_{1} e {\mathbb U}_{1} e - \kappa {\mathbb U}_{1} e
= e {\mathbb U}_{1} e {\mathbb U}_{1} - \kappa e {\mathbb
U}_{1}, \nonumber\\ \big [{\mathbb U}_{i}, e \big ] = 0, \qquad i \neq 1.
\label{blob}  \end{gather} We are free to renormalize $e$,
changing only $\delta_{e}$ and $\kappa$ (by the same factor). The
quantities~$\delta$,~$\delta_{e}$,~$\kappa$ are expressed in terms of
only two relevant parameters $q$ and ${\mathrm Q}$ associated to the
bulk and boundary parameters of the anticipated physical systems. It
is convenient to parametrize $\delta_{e}$ and~$\kappa$  as: \begin{gather}
\delta_{e} =-{\mathrm Q} - {\mathrm Q}^{-1}, \qquad \kappa =
q{\mathrm Q}^{-1} + q^{-1} {\mathrm Q}. \label{param}\end{gather} In fact
such a parametrization is quite natural from the point of view of
the cylinder braid group.

It is known \cite{jimbo} that any tensor space representation $\pi:
{\cal H}_{N}(q) \to \mbox{End}({\mathbb V}^{\otimes N})$ gives a
solution to the Yang--Baxter equation \cite{baxter,baxter1,korepin,korepin1}
\begin{gather}
 R_{12}(\lambda_{1}-\lambda_{2}) R_{13}(\lambda_{1})
R_{23}(\lambda_{2}) = R_{23}(\lambda_{2}) R_{13}(\lambda_{1})
R_{12}(\lambda_{1}-\lambda_{2}), \label{YBE}
\end{gather}
 written as 
 \begin{gather}
R_{12}(\lambda) ={\cal P}_{12} \big (\sinh \mu
(\lambda+{\mathrm i})  {\mathbb I} +\sinh (\mu\lambda) \pi
({\mathbb U}_{1}) \big ), \label{rr} 
\end{gather} where ${\cal P}$ is the
permutation operator on ${\mathbb V} \otimes {\mathbb V}$. Suppose
$\pi$ extends to a representation of ${\cal B}_N$. Then a solution
of the ref\/lection equation \cite{cherednik, sklyanin}
\begin{gather}
R_{12}(\lambda_{1} -\lambda_{2})\ K_{1}(\lambda_{1})
R_{21}(\lambda_{1}+\lambda_{2})\ K_{2}(\lambda_{2})\nonumber\\
\qquad{}=K_{2}(\lambda_{2}) R_{12}(\lambda_{1}+\lambda_{2})
K_{1}(\lambda_{1})  R_{21}(\lambda_{1} -\lambda_{2}) \label{RE}
\end{gather} can be written with the help of (\ref{TL}),
(\ref{blob})  (see also \cite{doikou2, ourpaper}) as 
\begin{gather}
K_{1}(\lambda) = x(\lambda) {\mathbb I} + y(\lambda) \pi(e),\label{ansatz1} 
\end{gather}
where  
\begin{gather}
x(\lambda)=
-\delta_{e}\cosh \mu (2\lambda +{\mathrm i}) - \kappa \cosh (2\mu
\lambda) -\cosh (2 {\mathrm i}\mu \zeta), \nonumber\\
 y(\lambda)=2
\sinh (2\mu\lambda) \sinh ({\mathrm i}\mu), \label{ansatz2} 
\end{gather} 
here
$\zeta$ is an arbitrary constant. By imposing further constraints on
the $B$ type Hecke algebra one obtains another quotient called
boundary Temperley--Lieb (blob) algebra $b_{N}(q, {\mathrm Q})$
-- an extension of the Temperley--Lieb algebra $T_{N}(q)$ -- def\/ined
by (\ref{TL}), (\ref{blob}) and also 
\begin{gather*}  {\mathbb U}_{i} {\mathbb
U}_{i+1} {\mathbb U}_{i}= {\mathbb U}_{i}, \qquad {\mathbb U}_{1}
e {\mathbb U}_{1} = \kappa {\mathbb U}_{1}. 
\end{gather*} The
expressions (\ref{rr}), (\ref{ansatz1}) are apparently valid for any
representation of the blob algebra as well. In what follows we shall
brief\/ly review two basic representations of the blob algebra, i.e.\
the XXZ and the asymmetric twin.

\subsection*{(I) The XXZ representation}

\label{cablingX}
\newcommand{\rrep}[1]{{\cal R}_{#1}}
\newcommand{\nm}{n}

For any given $N$ let the map $\rrep{q}$ on the generators of
$T_{N}(q)$ into $\mbox{End}({\mathbb V}^{\otimes N})$ be given by
\begin{gather} \rrep{q} ( {\mathbb U}_{l} )  = {\mathbb I} \otimes {\mathbb
I} \otimes \cdots \otimes U(q)
 \otimes \cdots \otimes {\mathbb I} \otimes {\mathbb I},
\label{tlg} 
\end{gather} where 
\begin{gather*} U(q)= \left( \begin{array}{cccc}
 0    &0        &0       &0   \\
 0    &-q       & 1      &0   \\
 0    &1        &-q^{-1} &0   \\
 0    &0        &0       &0
\end{array} \right)
\end{gather*}
(acting non-trivially  on ${\mathbb V}_{l} \otimes {\mathbb
V}_{l+1}$). This is the usual XXZ representation of the
Temperley--Lieb algebra. This may be extended to a representation of
$b_{N}(q, Q)$ by also introducing the matrix~\cite{masa} \begin{gather*} {\cal
M} = -{\delta_{e} \over Q+Q^{-1}} \left(
\begin{array}{cc}
 -Q^{-1}   &1          \\
 1    &-Q
\end{array} \right) \otimes  \cdots \otimes {\mathbb I} \otimes {\mathbb I} 
\end{gather*}
acting non trivially on ${\mathbb V}_{1}$. There exists a
representation ${\cal R}_{q}: b_{N}(q,Q) \to \mbox{End}({\mathbb
V}^{\otimes N})$ provided by (\ref{tlg}) and ${\cal R}_{q}(e) ={\cal
M}$ \cite{masa}. For this we have: \begin{gather*} \delta_{e} = -{Q+Q^{-1}\over
2{\mathrm i} \sinh ({\mathrm i} \mu)}, \qquad \kappa = { qQ^{-1}
+q^{-1}Q\over 2{\mathrm i} \sinh ({\mathrm i} \mu)}. 
\end{gather*}

\subsection*{(II) The asymmetric twin representation}

By combining two XXZ representations with dif\/ferent quantum
parameters one may construct a novel representation known as
asymmetric twin representation \cite{MartinWoodcock01pre, ourpaper,
our2}. Indeed set \begin{gather*} r={\mathrm i}\sqrt{{\mathrm i}q}, \qquad \hat
r=\sqrt{{\mathrm i}q} , 
\end{gather*} so that $ r \hat r = -q$ (${\mathrm i} =
\sqrt{-1}$). The representation $\Theta: T_{N}(q) \to
\mbox{End}({\mathbb V}^{ \otimes 2N})$ (introduced in
\cite{MartinWoodcock01pre}) is constructed by combining parts of the
representations $\rrep{r}$ of $T_{2N}(r)$ and $\rrep{\hat r}$ of
$T_{2N}(\hat r)$ as follows: 
\begin{gather*} \Theta({\mathbb U}_l) = \rrep{r} (
{\mathbb U}_{N-l} )  \rrep{\hat r} ( {\mathbb U}_{N+l} ).
\label{Theta}
\end{gather*} A striking feature of this
is that extends to a representation of $b_{N}$ in a variety of ways \cite{ourpaper, our2}:

Fixing $q$, def\/ine matrices in $\mbox{End}({\mathbb V}^{\otimes
2N})$ as follows: \begin{gather} {\cal M}^{i}(Q)=
 -{\delta_{e} \over Q+Q^{-1}} \;  {\mathbb I} \otimes {\mathbb I} \otimes \cdots
\otimes\left( \begin{array}{cccc}
 0   &0        &0       &0   \\
 0    &-Q      & 1     &0   \\
 0    &1        &-Q^{-1} &0   \\
 0   &0        &0       &0
\end{array} \right) \otimes \cdots \otimes {\mathbb I} \otimes {\mathbb I}, \label{cabl}
\end{gather} where the $4\times 4$ matrix acts on ${\mathbb V}_{N} \otimes
{\mathbb V}_{N+1}$ and $Q$ is some scalar; \begin{gather} {\cal M}^{ii}(Q) =
 -{\delta_{e} \over Q+Q^{-1}}  {\mathbb I} \otimes
{\mathbb I} \otimes \cdots \otimes\left( \begin{array}{cccc}
-Q    &0       &0      &1   \\
 0    &0       &0      &0   \\
 0    &0       &0      &0   \\
 1    &0       &0      &- Q^{-1}
\end{array} \right) \otimes \cdots \otimes {\mathbb I} \otimes {\mathbb I},
\label{cabl2} \\ {\cal M}^{+}(Q) = {\cal M}^{i}(Q) + {\cal
M}^{ii}(Q), \label{cable+} \\ {\cal M}^{iii}(Q_1,Q_2) =
 {\delta_{e} \over (Q_{1}+Q_{1}^{-1})(Q_{2}+Q_{2}^{-1})}   {\mathbb I} \otimes \cdots \otimes \left(
\begin{array}{cc}
 -Q_1    &1          \\
 1    &-Q_1^{-1}
\end{array} \right) \nonumber\\
\phantom{{\cal M}^{iii}(Q_1,Q_2) =}{} \otimes \left(
\begin{array}{cc}
 -Q_2^{-1}      &1          \\
 1    &-Q_2
\end{array} \right) \otimes \cdots  \otimes {\mathbb I},
 \label{cabl3}
 \end{gather}
  where the $2 \times 2$ matrices act
separately on ${\mathbb V}_{N}$, ${\mathbb V}_{N+1}$ respectively;
and 
\[
{\cal M}^{iii}(Q)={\cal M}^{iii}({\mathrm i}\sqrt{{\mathrm i}Q}
,  \sqrt{{\mathrm i}Q} ).
\]

For each $I \in \{i,ii,+,iii \}$ there is a representation $\Theta^I
: b_N \rightarrow \mbox{End}({\mathbb V}^{\otimes 2N})$ given by
$\Theta^I({\mathbb U}_i)=\Theta({\mathbb U}_i)$, $\Theta^I(e)={\cal
M}^I(Q)$, provided 
\begin{gather}
\delta_{e} = - {Q+Q^{-1} \over 2{\mathrm i}
\sinh ({\mathrm i}\mu)}, \qquad \kappa^{i} = {q^{-1}Q+qQ^{-1}\over 2
{\mathrm i} \sinh({\mathrm i}\mu)} , \qquad \kappa^{ii}= {{\mathrm
i} Q -{\mathrm i} Q^{-1} \over  2 {\mathrm i} \sinh ({\mathrm
i}\mu)},
\nonumber\\
\kappa^{+}=\kappa^{i}+\kappa^{ii} \qquad \kappa^{iii} =
{q^{-1}Q+qQ^{-1}+2\over 2 {\mathrm i} \sinh ({\mathrm i}\mu)}.
\label{param2} 
\end{gather}
 Notice that the algebraic parameter ${\mathrm Q}$
and the parameter $Q$ of the representation do not ne\-ces\-sarily
coincide. For the XXZ case and the representation (ii) they do, up
to appropriate normalization, but in general the one parameter may
be expressed in terms of the other by simply comparing
(\ref{param}), (\ref{param2}).

Let us f\/inally note that in \cite{our2} a novel class of
representations of the (boundary) Temperley--Lieb algebra, called
crossing representation, was introduced. In fact it was shown that
both the XXZ and the asymmetric twin representation belong to this
class. For a more detailed analysis of this type of representations
we refer the reader to \cite{our2}.

\section{The open spin chain}

Our objective is to examine the symmetry of the transfer matrix of
the asymmetric twin model. To achieve that we need to introduce the
$R$ and $K$ matrices associated to the asymmetric twin
representation of $T_{N}(q)$, and then build the corresponding open
transfer matrix \cite{sklyanin}. We shall also examine the open XXZ
chain, mainly to familiarize ourselves with the approach, therefore
we shall also recall the XXZ $R$ and $K$ matrices.

\subsection*{(I) The XXZ representation}

Let us f\/irst brief\/ly recall the form of the $R$ and $K$ matrices for
the XXZ representation. The XXZ $R$ matrix, acting on $\mathbb C^{2}
\otimes \mathbb C^{2}$ is given by 
\begin{gather*}
 R_{ l\, l+1}(\lambda)= \sinh
\mu(\lambda+{\mathrm i}) {\cal P}_{l\, l+1} + \sinh(\mu \lambda)
{\cal P}_{l\, l+1} {\cal R}_{q}({\mathbb U}_{l}), 
\end{gather*}
where  ${\cal P}$ is the permutation operator acting on $({\mathbb
C}^{2})^{\otimes 2}$, and ${\cal R}_q$ is the XXZ representation of
the Temperley--Lieb algebra given by (\ref{tlg}). The XXZ $K$ matrix
is given in a $2\times 2$ form as 
\begin{gather}
 K_1(\lambda) = x(\lambda)
{\mathbb I} + y(\lambda){\cal M}. \label{dv} 
\end{gather}
The latter matrix
coincides with the solution found in \cite{DVGR, GZ} subject to
certain identif\/ications among the various boundary parameters (for
more details see \cite{doikoun}).

\subsection*{(II) The asymmetric twin representation}

To derive the $R$ and $K$ matrices of the twin representation it is
convenient to introduce the following relabelling (see also
\cite{our2}), 
\begin{gather}
 N-l+1 \to l^-, \qquad N+l \to l^+, \label{fold}
 \end{gather}
then we can write 
\begin{gather*}
 \Theta({\mathbb U}_{l}) = U_{l^-(l+1)^-}(r^{-1}) U_{l^+ (l+1)^+}(\hat r), 
\end{gather*}
 acting on ${\mathbb V}_{\tilde l} \otimes {\mathbb V}_{\widetilde{l+1}}$, where the
index $\tilde l = (l^-, l^+)$ in the space/mirror space notation~\cite{our2}. 
The $16 \times 16$ explicit expression of
$\Theta({\mathbb U}_{l})$ is given in the Appendix A. The asymmetric
twin $R$ matrix is given by the following expression \cite{ourpaper,
our2} (the tilted indices are suppressed from $R$, $K$ from now on
for the sake of simplicity) 
\begin{gather*}
 R_{l\, l+1}(\lambda)= \sinh
\mu(\lambda+{\mathrm i}) {\cal P}_{ l\, l+1}+ \sinh(\mu \lambda)
{\cal P}_{l\, l+1} \Theta({\mathbb U}_{l}), 
\end{gather*} where now
the permutation operator ${\cal P}$ acts on $({\mathbb
C}^{4})^{\otimes 2}$. The asymmetric twin $K$ matrix can be written
in $4\times 4$ form with the help of the representations (i), (ii),
(iii) as 
\begin{gather}
 K_1^I(\lambda) =x(\lambda) {\mathbb I} + y(\lambda)
{\cal M}^{I}(Q), \label{K3} 
\end{gather}
 $x(\lambda)$, $y(\lambda)$ are given
by (\ref{ansatz2}), and $M^I$ given by (\ref{cabl})--(\ref{cabl3})
(for explicit $4 \times  4$ expressions see~\cite{our2}).

\subsection*{The transfer matrix}

Given any $R(\lambda) \in \mbox{End}({\mathbb V}\otimes {\mathbb
V})$ one may def\/ine the more general object $L(\lambda) \in
\mbox{End}({\mathbb V})\otimes {\cal A} $ where ${\cal A}$ is the
algebra associated to the $R$ matrix and def\/ined by the following
fundamental relation 
\begin{gather} R_{12}(\lambda_{1} -\lambda_{2})
L_{1}(\lambda_{1}) L_{2}(\lambda_{2}) = L_{2}(\lambda_{2})
L_{1}(\lambda_{1}) R_{12}(\lambda_{1} -\lambda_{2}). \label{funda}
\end{gather} It is clear that when the second space, which is a copy of
${\cal A}$, is mapped to ${\mathbb V}$ then $L(\lambda) \mapsto 
R(\lambda)$. Note that for the XXZ case the $L$ matrix is known and
${\cal A} = {\cal U}_q(\widehat{sl_{2}})$ , whereas for the
asymmetric twin model a generic $L$ matrix is not available for the
moment. Hence whenever we discuss about the asymmetric twin case we
restrict ourselves to the $R$ matrix only.

In general given any $L$ and $K$ matrices one can derive the
algebraic open transfer matrix, which provides the conserved
quantities of the model. Def\/ine f\/irst the monodromy matrix
$T(\lambda)$, ($\hat T(\lambda) = T^{-1}(-\lambda)$) \cite{faddeev}
\begin{gather*}
 T_{ 0}(\lambda) = L_{0 N}(\lambda) \cdots L_{02}(\lambda) L_{ 0
1}(\lambda), \qquad \hat T_{0}(\lambda) = \hat L_{ 0 1}(\lambda) \hat
L_{02}(\lambda) \cdots \hat L_{0N}(\lambda), 
\end{gather*}
 where
$\hat L(\lambda)= L^{-1}(-\lambda)$. The monodromy matrix satisf\/ies
fundamental algebraic relation (\ref{funda}) as well. The transfer
matrix of the open chain \cite{sklyanin} is def\/ined as 
\begin{gather}
t(\lambda)= {\rm tr}_{ 0} \big \{ M_{ 0}\ K_{ 0}^{(L)}(\lambda) {\cal
T}_{ 0}(\lambda) \big \},\qquad {\cal T}_{ 0}(\lambda)= T_{
0}(\lambda) K_{0}^{(R)}(\lambda) \hat T_{0}(\lambda).
\label{transfer} 
\end{gather} Note that as customary the `quantum' indices
$1, \ldots, N$ are suppressed from $T$, $\hat T$ and ${\cal T}$.
Recall that for the moment the quantum spaces are not represented,
but they are simply copies of the algebra ${\cal A}$, hence the
transfer matrix at this stage is a purely algebraic object
$t(\lambda)\in {\cal A}^{\otimes N}$. The operator ${\cal T}$ is
also a solution of the ref\/lection equation, $K^{(L)}$ is associated
to the left boundary of the spin chain, and in what follows it will
be unit, and $K^{(R)}$ denotes the right boundary of the chain. It
will be either unit, or given by (\ref{dv}) for XXZ, and by
(\ref{K3}) for the asymmetric twin model. The matrix $M$ is given by
\begin{alignat}{3}
& \mbox{XXZ:} \quad && M ={\rm diag}\,(q, q^{-1}), \nonumber\\
& \mbox{Twin:} \quad && M= {\rm diag}\,({\mathrm i},  q^{-1}, q, -{\mathrm i}), & \label{m1}
\end{alignat} where recall $q=-r \hat r$.

As mentioned our main aim is to study the symmetry of the twin
transfer matrix. Usually this is carried out by investigating the
asymptotics of ${\cal T}$ as $\lambda \to \infty $ (see e.g.\
\cite{done,done1,done2}). However, as discussed in \cite{our2} the
study of the asymptotics for the twin $R$ matrix and consequently
for ${\cal T}$ is a rather intricate task. Hence one can not easily
obtain recognized conserved non-local charges with a simple
coproduct structure, as is the case in e.g.\ the XXZ model. To
circumvent this complication we adopt the approach developed in
\cite{doikoun, doikou2}, and will be reviewed in the Section~5. More
precisely using the methodology of \cite{doikoun, doikou2} we shall
be able to identify familiar non-local charges, which however do not
provide the full symmetry of the asymmetric twin model, as already
pointed out in~\cite{our2}. Nevertheless in the subsequent section
we shall give a generic description of the symmetry algebra emerging
in any lattice model with particular integrable boundaries.

\section{The underlying algebra and the symmetry}

As argued in \cite{sklyanin} given any $L$, $K$ matrices, solutions
of the fundamental relation (\ref{funda}) and ref\/lection equation
(\ref{RE}) respectively, one may build the more general solution of
(\ref{RE}) 
\begin{gather} {\mathbb K}(\lambda' \mp \lambda)= L(\lambda' \mp
\lambda) (K(\lambda')\otimes {\mathbb I})  \hat L(\lambda'
\pm\lambda), \label{gensol} 
\end{gather} where $K$ is a c-number solution of
the ref\/lection equation. The entries of the matrix ${\mathbb K}$ are
elements of the so called ref\/lection algebra ${\mathbb R}$~(\ref{RE}) (see also~\cite{sklyanin}).

One may easily show that all the elements of the ref\/lection algebra
`commute' with the solutions of the ref\/lection equation (see also
\cite{dema}). Indeed recalling (\ref{RE}) and the above 
expres\-sions~(\ref{gensol}) it is straightforward to show that 
\begin{gather} {\mathbb
K}_{ab}(\lambda'-\lambda)\ K(\lambda) =  K(\lambda)\ {\mathbb
K}_{ab}(\lambda'+\lambda), \label{bcomm} 
\end{gather} which implies that any
solution of the ref\/lection equation commutes with the elements of
the ref\/lection algebra and the opposite.

The ref\/lection algebra is also endowed with a coproduct. More
precisely, the $L$ matrix is equipped with a coproduct derived from
the Yang--Baxter equation $\Delta: {\cal A} \to {\cal A} \otimes
{\cal A}$, where recall ${\cal A}$ is the algebra def\/ined by
(\ref{funda}) (see also \cite{jimbo}) 
\begin{gather}
 \Delta(L_{ab}) = \sum_{c}
L_{cb} \otimes L_{ac}, \qquad \hat L_{ab} =\sum_{c} \hat L_{ac}
\otimes \hat L_{cb}. \label{dd} 
\end{gather} $L_{ab}$ are the entries of the
$L$ matrix. It is also convenient to introduce $\Delta' \ = \sigma
\circ \Delta$ where $\sigma: a \otimes b \to b\otimes a$. The $n$
coproduct is obtained by iteration: $ \Delta^{(n)} = (\mbox{id}
\otimes \Delta^{(n-1)})\Delta$, $\Delta^{'(n)} = (\mbox{id} \otimes
\Delta^{(n-1)})\Delta'$.

Taking into account (\ref{dd}) we conclude that the ref\/lection
algebra is also endowed with a~coproduct, $\Delta: {\mathbb R} \to
{\mathbb R} \otimes {\cal A}$ (see also \cite{dema})
\begin{gather*}
\Delta({\mathbb K}_{ab}(\lambda)) = \sum_{k,l=1}^n{\mathbb
K}_{kl}(\lambda) \otimes L_{ak}(\lambda) \hat L_{lb}(\lambda).
\end{gather*}
 In fact, the entries ${\cal T}_{ab}$ are simply $N$
coproducts of the elements of the ref\/lection algebra (\ref{RE}),
i.e.\ ${\cal T}_{ab}(\lambda)= \Delta^{(N)}({\mathbb
K}_{ab}(\lambda))$. It may be shown, bearing in mind the fact that
${\cal T}$ is also a solution of the (\ref{RE}), that (see also
\cite{doikoun, doikou2}) 
\begin{gather} \Delta'^{(N+1)}({\mathbb
K}_{ab}(\lambda'-\lambda))\ {\cal T}(\lambda) = {\cal T}(\lambda)
\Delta'^{(N+1)}({\mathbb K}_{ab}(\lambda'+\lambda)). \label{it0} 
\end{gather}
The generic solution (\ref{gensol}) allows the asymptotic expansion
as $\lambda' \to \pm \infty$ providing the ref\/lection algebra
generators. The f\/irst order of such expansion yields the generators
of the boundary quantum algebra ${\mathbb B}$, which obey exchange
relations dictated by the def\/ining relations (\ref{RE}) as $\lambda'
\to \infty$. More precisely, as $\lambda' \to \pm \infty$
$~R(\lambda') \propto R^{\pm} $, $K(\lambda')\to K^{\pm}$ and
${\mathbb K} \to {\mathbb K}^{\pm}$, where $R^{\pm}$, $K^{\pm}$,
${\mathbb K}^{\pm}$ have no spectral dependance anymore,
\begin{gather} R^{\pm(\mp)}_{12}\ {\mathbb K}_{1}^{\pm} \hat R_{12}^{\pm}
{\mathbb K}_{2}^{\pm} = {\mathbb K}_{2}^{\pm}\  R_{12}^{\pm}
{\mathbb K}_{1}^{\pm} \hat R_{12}^{\pm(\mp)} \label{rr2}
\end{gather} and the entries ${\mathbb K}^{\pm}_{ab}$ form
the boundary quantum algebra ${\mathbb B}$. The later formula
(\ref{rr2}) is in fact an~immediate consequence of the quadratic
relation of the $B$ type Hecke algebraic (cylinder braid group)
(\ref{blob}) (for further comments on this point see also
\cite{doikou2}).

Let us stress once more that depending on the choice of the $R$
matrix one obtains distinct (boundary) quantum algebras. Moreover,
the explicit form of the boundary quantum algebra generators depends
on the choice of $R$ matrix as well as the choice of $K$ matrix. For
instance for the XXZ case with $K={\mathbb I}$ the corresponding
boundary quantum algebra coincides with ${\cal U}_{q}(sl_{2})$
\cite{kusk}, whereas if $K$ is given by (\ref{dv}) the boundary
quantum algebra consists of only one element (abelian), which may be
expressed in terms of the ${\cal U}_{q}(sl_{2})$ generators (see
later in the text in Section~5, and also in \cite{doikoun}). On the
other hand, for models associated to higher rank algebras for
trivial boundary conditions the symmetry of the transfer matrix
coincides with ${\cal U}_q(gl_n)$ ($K^{(L,R)} ={\mathbb I})$ or
${\cal U}_q(gl_l) \otimes {\cal U}_q(gl_{n-l})$ ($K^{(L)} ={\mathbb
I}$, $K^{(R)} ={\rm diag}$) \cite{done2}. For a particular non diagonal
right boundary on the other hand the boundary non-local charges form
a non-Abelian boundary quantum algebra identif\/ied in \cite{doikou2}.
Exploiting (\ref{it0}) for $\lambda' \to \pm \infty$, i.e. 
\begin{gather}
\sum_{b, c} R_{ab}^{\pm} \hat R_{cd}^{\pm} \otimes {\cal
T}_{bc}^{\pm} {\cal T}(\lambda) = {\cal T}(\lambda)  \sum_{b, c}
R_{ab}^{\pm} \hat R_{cd}^{\pm} \otimes {\cal T}_{bc}^{\pm}
\label{git} 
\end{gather}
 one could show that the `boundary non local charges'
${\cal T}_{ab}^{\pm} =\Delta^{(N)}({\mathbb K}^{\pm}_{ab})$ commute
with the asymmetric twin transfer matrix \cite{doikoun, doikou2}. We
do not attempt this rather technical proof here, although it is a
routine exercise based on (\ref{git}) and the explicit expression of
the $R^{\pm}$ matrices (see e.g.\ \cite{doikoun, doikou2} for the XXZ
and ${\cal U}_{q}(gl_{n})$ cases).

$\bullet$ {\it Symmetry:} Instead we shall show that the boundary
quantum algebra ${\mathbb B}$ is a symmetry for any open transfer
matrix with trivial left boundary following a dif\/ferent approach
(see also \cite{done2}). Consider any open transfer matrix
(\ref{transfer}) with trivial left boundary $K^{(L)}= {\mathbb I}$,
$K^{(R)}(\lambda) \in \mbox{End}({\mathbb C^n})$ is any solution of
the ref\/lection equation of the type (\ref{ansatz1}), and $R(\lambda)
\in \mbox{End}({\mathbb C}^{n} \otimes {\mathbb C}^{n})$ is a
solution of the Yang--Baxter equation (\ref{rr}). Assume also that
the $R$ matrix satisf\/ies the following: 
\begin{gather}
 R_{12}(\lambda)
R_{21}(-\lambda) \propto {\mathbb I}, \qquad M_{1}^{-1}
R_{12}^{t_{1}}(\lambda) M_{1} R_{21}^{t_{1}}(-\lambda -2{\mathrm
i} \rho) \propto {\mathbb I}, \label{cond} 
\end{gather}
 $\rho$ is a constant
and for both XXZ and the asymmetric twin representations is unit
(for the ${\cal U}_q(gl_n)$ case $\rho = {n \over 2}$). Suppose also
that the matrix $M$ satisf\/ies: 
\begin{gather}
 M =M^t, \qquad \big [M_{1}
M_{2}, R_{12}(\lambda)\big ] =0, \label{cond2} 
\end{gather} and for the XXZ
and asymmetric twin models is given by (\ref{m1}). We also introduce
the following quantity, which structurally resembles the open
transfer matrix \cite{done2}, i.e. 
\begin{gather*}
 \tau^{\pm }= tr_{0}\big
\{{\mathbb P}_{0} {\cal T}_0^{\pm} \big \}, \qquad {\cal T}^{\pm}
\propto {\cal T} (\lambda \to \pm \infty), 
\end{gather*} with
${\mathbb P}$ being a priori an arbitrary $n \times n$ matrix. Using
(\ref{cond}), (\ref{cond2}) and the fact that ${\cal T}$ satisf\/ies
the ref\/lection equation (\ref{RE}), in particular we employ
(\ref{RE}) for $\lambda_{1} \to \pm \infty$, we may show following
the steps of the proof of integrability in \cite{sklyanin} (for more
details see e.g.~\cite{done2}) that 
\begin{gather*} \big [ \tau^{\pm},
t(\lambda) \big ] =0. 
\end{gather*}
 By choosing the matrix
${\mathbb P}$ appropriately, i.e.\ ${\mathbb P}=E_{ab}$ where we
def\/ine $(E_{ab})_{cd} = \delta_{ac} \delta_{bd}$, it is clear that
\begin{gather}
 \tau^{\pm}= {\cal T}_{ab}^{\pm} \quad \Rightarrow \quad \big [ {\cal
T}_{ab}^{\pm},\ t(\lambda) \big ] =0, \label{symme2} 
\end{gather} and this
concludes our proof.

Consider now a non trivial left boundary $K^{(L)} \neq {\mathbb I}$.
The proof of the corresponding symmetry goes along the lines
described above although it is technically more involved and it will
be reported elsewhere. The intricate point in this case is that one
has to impose a series of requirements satisf\/ied by $R$, $K^{(L,R)}$ 
and ${\mathbb P}$ matrices. In any case the corresponding
conserved quantities are linear combinations of ${\cal
T}_{ab}^{\pm}$, hence the remaining symmetry is a consistent subset
of the boundary quantum algebra ${\mathbb B}$.

$\bullet$ {\it Centralizer:} We shall now restrict our attention to
the case where both quantum and auxiliary spaces correspond to
${\mathbb V}$, thus $L \mapsto R$ and the open transfer matrix is
not an algebraic object any more as in (\ref{transfer}), but is
mapped to $\mbox{End}({\mathbb V}^{\otimes N})$, 
\begin{gather} t(\lambda)
={\rm tr}_{0}\big \{M_{0}\ K_0^{(L)}(\lambda) {\cal T}_0(\lambda) \big
\}, \nonumber\\ 
 {\cal T}_0(\lambda)= R_{0N}(\lambda) \cdots
R_{01}(\lambda) K_0^{(R)}(\lambda) \hat R_{01}(\lambda) \cdots
\hat R_{0N}(\lambda).  \label{transfer1} 
\end{gather}
 Due to the fact that
the $R$ matrix reduces to the permutation operator ${\cal P}$ for
$\lambda =0$, a local Hamiltonian may be deduced from
(\ref{transfer1}) (${\cal H} \propto {d t(\lambda) \over d
\lambda}|_{\lambda =0}$), which can be expressed in terms of
representations of the $B$-type Hecke algebra (for more details see
e.g.~\cite{doikou2, our2}).

The boundary quantum algebra is also represented, and we shall show
that apart from providing a symmetry of the open transfer matrix
(\ref{transfer1}) with trivial left boundary, is also the
centralizer of the $B$ type Hecke algebra. Indeed consider solutions
of the Yang--Baxter and ref\/lection equations expressed as in
(\ref{rr}), (\ref{ansatz1}). Set $\check R ={\cal P} R$, also
consider the parametrization~(\ref{param}), also recall that $\pi:
{\cal B}_{N} \to \mbox{End}({\mathbb V}^{\otimes N})$ then from
(\ref{rr}), (\ref{ansatz1}) as $\lambda \to \pm \infty$ it follows
that 
\begin{gather} \check R_{i\ i+1}(\lambda \to \pm \infty) \propto \check
R_{i\ i+1}^{\pm} = \pi({\mathbb U}_{i}) + q^{\pm 1},
\qquad K_{1}(\lambda \to \pm \infty) \propto K_1^{\pm} = \pi(e)
+{\mathrm Q}^{\pm 1}. \label{asy} 
\end{gather} From the Yang--Baxter equation
(\ref{YBE}) for $\lambda_{1,2} \to \pm \infty$ one obtains 
\begin{gather}
\check R^{\pm}_{i\, i+1} R^{\pm}_{0\, i+1}\ R^{\pm}_{0i}=R^{\pm}_{0\,
i+1}\ R^{\pm}_{0i} \check R^{\pm}_{i\ i+1}, \qquad \check
R^{\pm}_{i\, i+1} \hat R^{\pm}_{0 i } \hat R^{\pm}_{0\, i+1}= \hat
R^{\pm}_{0 i} \hat R^{\pm}_{0\, i+1} \check R^{\pm}_{i\, i+1}.
\label{braid} 
\end{gather} Finally recalling the structure of ${\cal T}$
(\ref{transfer1}), and using relations (\ref{asy}), (\ref{braid})
and the ref\/lection equation we obtain 
\begin{gather}
 \big [\pi({\mathbb
U}_{i}), {\cal T}^{\pm}_{ab} \big ]=0, \qquad \big [\pi(e), {\cal
T}^{\pm}_{ab} \big ]=0,  \qquad i\in \{1, \dots,  N-1 \}, \label{a}
\end{gather}
 where now apparently ${\cal T}_{ab}^{\pm} \in
\mbox{End}({\mathbb V}^{\otimes N})$. The latter commutation
relations imply the duality between the boundary quantum algebra and
the $B$ type Hecke algebra (for a relevant discussion see also
\cite{sap}). When dealing with representations of the blob algebra
such as the XXZ and the asymmetric twin, the boundary quantum
algebra becomes evidently the centralizer of the blob algebra
\cite{doikoun}.

\section{Familiar conserved quantities}

The presentation of the previous section relies primarily on
abstract algebraic considerations, and as such it does not of\/fer
explicit expressions of the algebra generators, and consequently of
conserved quantities that determine the symmetry of the open spin
chain. Usually the study of the symmetry of an open transfer matrix
(see e.g.~\cite{done, done1, done2, kusk}) rests on the fact that
the monodromy matrix $T$ reduces to upper (lower) triangular matrix
as $\lambda \to \pm \infty$, which facilitates enormously the
algebraic manipulations. There exist however cases such as the
asymmetric twin model, where the monodromy matrix does not reduce to
such a convenient form as $\lambda \to \pm \infty$. In these cases
the most ef\/fective way to investigate the corresponding symmetry is
to derive in some way (e.g.~by direct computation) linear
intertwining relations of the type (\ref{bcomm}) and~(\ref{it0}), by
means of which exchange relations between the entries of ${\cal T}$
and the corresponding non-local charges can be deduced. This is
carried out in what follows, where we provide explicit expressions
of conserved quantities associated to familiar quantum algebras.

Before we proceed with the derivation of recognized conserved
quantities for the asymmetric twin model it is useful to recall
basic def\/initions regarding the quantum algebra ${\cal U}_{{\mathrm
q}}(sl_2)$. Let ${\cal E}$, ${\cal F}$ and ${\cal H}$ be the
generators of the quantum algebra ${\cal U}_{{\mathrm q}}(sl_2)$
\cite{jimbo, drinf}, satisfying the def\/ining relations, 
\begin{gather*}
 \big
[{\cal E}, {\cal F}\big ] = {{\cal H}^{2} -{\cal H}^{-2} \over
{\mathrm q}-{\mathrm q}^{-1}} , \qquad {\cal H} {\cal E} = {\mathrm
q}\ {\cal E} {\cal H}, \qquad {\cal H}\ {\cal F} = {\mathrm q}^{- 1}
{\cal F} {\cal H}.
\end{gather*}
 There is a coproduct $\Delta$:
${\cal U}_{{\mathrm q}}(sl_2) \to {\cal U}_{{\mathrm q}}(sl_2)
\otimes {\cal U}_{{\mathrm q}}(sl_2)$ given by 
\begin{gather*}
\Delta(\chi) =
{\cal H}^{-1} \otimes \chi + \chi \otimes {\cal H},\qquad \chi \in
\{{\cal E}, {\cal F} \}, \qquad \Delta({\cal H}^{\pm 1}) = {\cal
H}^{\pm 1} \otimes {\cal H}^{\pm 1} . 
\end{gather*}
 The $n$
coproducts are obtained by iteration as def\/ined in the previous
section.

Henceforth we shall focus on the case where both spaces quantum and
auxiliary are associated to ${\mathbb V}$, we shall deal basically
with the transfer matrix (\ref{transfer1}). Let us make a general
statement, which shall be used for both the XXZ and the asymmetric
twin models. Consider a generic $R$ matrix, solution of the
Yang--Baxter equation, satisfying the following intertwining
relations with a representation 
${\mathrm h}: {\cal U}_q(sl_2) \to \mbox{End}({\mathbb V})$
\begin{gather}
 {\mathrm h}^{\otimes
2}(\Delta'(x)) R_{12}(\lambda) =R_{12}(\lambda) {\mathrm
h}^{\otimes 2}(\Delta(x)), \qquad x \in \{ {\cal E}, {\cal F}, {\cal
H} \}. \label{ir} 
\end{gather}
 Then one can show in a straightforward fashion
that generalized intertwining relations hold also for the operator
${\cal T}$ (\ref{transfer1}) for $K^{(L, R)} ={\mathbb I}$ (see
also \cite{doikoun, doikou2} for a detailed proof) 
\begin{gather}
 ({\mathrm h}
\otimes {\mathrm h}^{\otimes N}) \Delta^{'(N+1)}(x) {\cal
T}(\lambda) = {\cal T}(\lambda) ({\mathrm h} \otimes {\mathrm
h}^{\otimes N})\Delta^{'(N+1)}(x). \label{itt} 
\end{gather}
 The latter
relations are exactly of the type (\ref{it0}) with no spectral
parameter, i.e.\ for $\lambda' \to \pm \infty$. Although similar
relations hold for the generators of the af\/f\/ine ${\cal U}_{{\mathrm
q}}(\widehat{sl_{2}})$ algebra, here we restrict our attention to
the non af\/f\/ine case ${\cal U}_{{\mathrm q}}(sl_{2})$. The reason for
such a restriction is the fact that the asymmetric twin $R$ matrix,
which is our main concern, does not satisfy any obvious intertwining
relations with the elements of ${\cal U}_{{\mathrm
q}}(\widehat{sl_2})$. As a consequence no generalized intertwining
relations can be derived, and that is why we remain focused on the
non-af\/f\/ine case.

It will be useful for the remaining part to present the operator
${\cal T}$ (\ref{transfer1}) in a matrix form for both the XXZ and
the asymmetric twin representations respectively
\begin{gather}
 {\cal
T}(\lambda) = \left(
\begin{array}{cc}
 {\cal A} &  {\cal B}\\
 {\cal C} &  {\cal D}   \\
\end{array} \right) \qquad \mbox{and} \qquad t(\lambda) =q{\cal A}+q^{-1}{\cal D}, \label{tr2}\\
{\cal T}(\lambda) = \left(
\begin{array}{cccc}
{\cal A}     &{\cal B}_{1} &{\cal B}_{2} &{\cal B}\\
{\cal C}_{1} &{\cal A}_{1} &{\cal B}_{5} &{\cal B}_{3}\\
{\cal C}_{2} &{\cal C}_{5} &{\cal A}_{2} &{\cal B}_{4}\\
{\cal C}     &{\cal C}_{3} &{\cal C}_{4} &{\cal D}\\
\end{array} \right)  \qquad \mbox{and} \qquad t(\lambda) ={\mathrm i}
{\cal A}+q^{-1}{\cal A}_{1}+q {\cal A}_{2} -{\mathrm i}{\cal D}.
\label{mono1} 
\end{gather}

\subsection[The XXZ open transfer matrix ($K^{(L, R)} ={\mathbb I}$)]{The XXZ open transfer matrix 
($\boldsymbol{K^{(L, R)} ={\mathbb I}}$)}

Let us f\/irst consider, mostly as a warm up exercise, the XXZ
representation. The case with trivial boundaries ($K^{(L, R)}
={\mathbb I}$) will be considered here. The symmetry of this model
has been already studied in \cite{kusk} extending the results of
\cite{pasa}, however here we rederive the result using the method
described in \cite{doikoun}.

Consider the representation $\rho: {\cal U}_{q}(sl_2) \to
\mbox{End}(\mathbb C^{2})$ def\/ined by 
\begin{gather*}
 \rho({\cal H})=q^{{1\over
2} \sigma^{z}}, \qquad \rho({\cal E})=\sigma^{+}, \qquad \rho({\cal
F})=\sigma^{-}, 
\end{gather*}
 where the parameter $q
=e^{{\mathrm i}\mu}$ coincides with the parameter of the
Temperley--Lieb algebra. It is convenient to introduce some
notations. Let {\samepage
\begin{gather}
 E^{(N)} = \rho^{\otimes N}(\Delta^{(N)}({\cal
E})), \qquad F^{(N)} = \rho^{\otimes N}(\Delta^{(N)}({\cal F})),
\qquad H^{(N)}= \rho^{\otimes N}(\Delta^{(N)}({\cal H})).
\label{delta}
\end{gather}
 It is clear that $E^{(N)}$, $F^{(N)}$, $H^{(N)}$
form a tensor representation of ${\cal U}_{q}(sl_2)$ acting on
$({\mathbb C}^{2})^{\otimes N}$.}

One can show that the $T_{N}(q)$ generators in the XXZ
representation (\ref{tlg}) commute with  the action of the quantum
group (\ref{delta}) (see also  \cite{our2, pasa}) 
\begin{gather}
 \big [
E^{(N)}, {\cal R}_{q}({\mathbb U}_{l})\big ] =\big [ F^{(N)},
{\cal R}_{q}({\mathbb U}_{l})\big ] = \big [H^{(N)}, {\cal
R}_{q}({\mathbb U}_{l}) \big ] =0, \qquad l \in \{1,\ldots, N-1\}.
\label{comre1} 
\end{gather}
 The commutation relations (\ref{comre1}) were
exploited in \cite{pasa} for proving that the Hamiltonian of the
open XXZ spin chain, with trivial boundaries $K^{(L, R)}={\mathbb
I}$, is ${\cal U}_{q}(sl_2)$ invariant.

The XXZ $R$ matrix \cite{jimbo2} satisf\/ies linear intertwining
relations (\ref{ir}) with the representation 
$\rho: {\cal U}_q(sl_2) \to \mbox{End}({\mathbb C}^2)$, and as a consequence the operator
${\cal T}$ for the XXZ model, with $K^{(L, R)} = {\mathbb I}$,
satisf\/ies (\ref{itt}).  From the generalized relations (\ref{itt})
and the form of the ${\cal T}$ matrix for the XXZ model (\ref{tr2})
algebraic Bethe ansatz type relations are entailed \cite{doikoun}
(see Appendix B for explicit expressions (\ref{co1})--(\ref{co3})).
With the help of relations (\ref{co1})--(\ref{co3}) it is possible
to study the symmetry of the open  transfer matrix. Indeed recall
(\ref{tr2}) then by virtue of the aforementioned relations it can be
easily shown that 
\begin{gather}
\big [ E^{(N)}, t(\lambda)\big ] =  \big [
F^{(N)}, t(\lambda)\big ] =\big [ H^{(N)}, t(\lambda)\big ] = 0.
\label{com0} 
\end{gather}
 The commutation relations between the transfer
matrix and the representations of the ${\cal U}_{q}(sl_2)$
generators (\ref{com0}) imply that the full transfer matrix enjoys
${\cal U}_{q}(sl_2)$ symmetry as already proved in~\cite{kusk}. Note
that for the f\/inite XXZ chain with periodic (or twisted) boundary
condition no symmetry has been identif\/ied for generic values of $q$.
However in the case where $q$ is root of unity it was shown in~\cite{demc} 
that the periodic XXZ chain enjoys the $sl_2$ loop
symmetry. The symmetry of the transfer matrix for $K^{(R)}$
non-diagonal given by (\ref{dv}) was studied in \cite{doikoun,
doikou2}, where it was shown that the transfer matrix of the XXZ
chain with non-diagonal right boundary commutes with the generator
of the boundary quantum algebra ${\mathbb B}({\cal U}_{q}(sl_2))$
\cite{mene, dema, doikoun}.

Let us mention that for the XXZ case the associated $L$ matrix is
known, and it satisf\/ies intertwining relations of the type
(\ref{ir}) (see e.g.\ \cite{jimbo2}) 
\begin{gather*}
 (\rho\otimes \mbox{id})
\Delta'(x) L_{12}(\lambda) =L_{12}(\lambda) (\rho \otimes
 \mbox{id})\Delta(x) \quad \Rightarrow \nonumber\\ 
  (\rho \otimes \mbox{id}^{\otimes
N}) \Delta^{'(N+1)}(x) {\cal T}(\lambda) = {\cal T}(\lambda)\ (\rho
\otimes \mbox{id}^{\otimes N})\Delta^{'(N+1)}(x),\qquad x \in \{ {\cal
E}, {\cal F}, {\cal H} \}. 
\end{gather*}
Hence one can show the
symmetry of the algebraic transfer matrix (\ref{transfer}). In fact
commutation relations (\ref{com0}) hold for the algebraic transfer
matrix (\ref{transfer}), but now $E^{(N)}$, $F^{(N)}$, $H^{(N)}$ are
not represented, namely 
\begin{gather*}
 E^{(N)} = \Delta^{(N)}({\cal E}), \qquad
F^{(N)} = \Delta^{(N)}({\cal F}), \qquad H^{(N)}= \Delta^{(N)}({\cal
H}). 
\end{gather*}

\subsection[The asymmetric twin open transfer matrix, $K^{(L,R)} ={\mathbb I}$]{The asymmetric 
twin open transfer matrix, $\boldsymbol{K^{(L,R)} ={\mathbb I}}$}

We come now to our main objective, which is the derivation of
conserved quantities, associated to familiar quantum algebras, for
the asymmetric twin transfer matrix. Unfortunately an $L$ matrix
associated to the asymmetric twin $R$ matrix is not available at
this stage, hence we are compelled to restrict ourselves to the
symmetry of the represented transfer matrix (\ref{transfer1}). We
shall f\/irst examine the case where both boundaries are trivial, i.e.\
$K^{(L,\ R)} ={\mathbb I}$.

$\bullet$ {\it The ${\cal U}_{q}(sl_2)$, ${\cal U}_{{\mathrm
i}}(sl_2)$ symmetries:} We introduce representations of ${\cal
U}_{{\mathrm i}}(sl_2)$, ${\cal U}_{q}(sl_2)$ respectively (recall,
$q=e^{i\mu}$
is a the parameter of the blob algebra and ${\mathrm i} = \sqrt{-1}$ f\/ixed):
\[
\sigma_{1}: {\cal U}_{{\mathrm i}}(sl_2) \to \mbox{End}(\mathbb
C^{2} \otimes \mathbb C^{2})\qquad {\rm and}\qquad \sigma_{2}: {\cal U}_{q}(sl_2)
\to \mbox{End}(\mathbb C^{2} \otimes \mathbb C^{2})
\]
 such that 
 \begin{gather}
\sigma_{1}({\cal H})={\mathrm i}^{{1\over 2}(e_{11} - e_{44})},
\qquad \sigma_{1}({\cal E}) = e_{14}, \qquad \sigma_{1}({\cal F}) =
e_{41},\nonumber\\ \sigma_{2}({\cal H})= q^{-{1\over 2}
(e_{22}-e_{33})},\qquad \sigma_{2}({\cal E}) =e_{32}, \qquad \sigma_{2}
({\cal F}) =e_{23}, \label{gen1} 
\end{gather}
 where $(e_{ij})_{kl}
=\delta_{ik} \delta_{jl}$. Let also
 \begin{gather*}
  E_{j}^{(N)} =
\sigma_{j}^{\otimes N}(\Delta^{(N)}({\cal E})) ,
\qquad F_{j}^{(N)}=\sigma_{j}^{\otimes N}(\Delta^{(N)}({\cal F})) ,\nonumber\\
H_{j}^{(N)}=\sigma_{j}^{\otimes n}(\Delta^{(N)}({\cal H})),
\qquad  j\in \{1, 2 \}. 
\end{gather*}
 It was shown in
\cite{our2} that \begin{gather}
 \big [E^{(N)}_{j}, \Theta({\mathbb U}_{l})
\big ] = \big [F^{(N)}_{j}, \Theta({\mathbb U}_{l}) \big ] =  \big
[H^{(N)}_{j}, \Theta({\mathbb U}_{l}) \big ] =0, \qquad l \in \{ 1,
\ldots, N-1\}.\label{2a}
\end{gather} $E^{(N)}_{1}$, $F^{(N)}_{1}$,
$H^{(N)}_{1}$ provide a tensor representations of ${\cal
U}_{{\mathrm i}}(sl_2)$ acting on $({\mathbb C}^{4})^{\otimes N}$,
while $E^{(N)}_{2}$, $F^{(N)}_{2}$, $H^{(N)}_{2}$ form a
representation of ${\cal U}_{q}(sl_2)$ acting on $({\mathbb
C}^{4})^{\otimes N}$. The commutation relations (\ref{2a}) were
exploited in~\cite{our2} to show that the Hamiltonian of the model
with $K^{(L, R)}={\mathbb I}$ is ${\cal U}_{q}(sl_2) \otimes {\cal
U}_{{\mathrm i}}(sl_2)$ symmetric.

We shall now show that the full transfer matrix (\ref{transfer1})
enjoys also the  ${\cal U}_{q}(sl_2) \otimes {\cal U}_{{\mathrm
i}}(sl_2)$ symmetry for $K^{(L, R)}={\mathbb I}$. To achieve that
we use the generalized intertwining relations between the operator
${\cal T}$ and the co-products of the quantum algebras ${\cal
U}_{q}(sl_2)$, ${\cal U}_{{\mathrm i}}(sl_2)$. As shown in~\cite{our2} 
the asymmetric twin $R$ matrix satisf\/ies intertwining
relations for both $\sigma_{1}$ and  $\sigma_{2}$, and consequently
generalized `commutation' relations are valid for the operator
${\cal T}$ of the asymmetric twin model~(\ref{transfer1}). By
exploiting these relations (\ref{itt}) for $\sigma_{j}$, and
recalling the form of ${\cal T}$ (\ref{mono1}) for the asymmetric
twin model exchange relations involving the entries of the ${\cal
T}$ operator (\ref{mono1}) are obtained. In Appendix B we report
only the necessary ones for the study of the transfer matrix
symmetry, although they are already quite involved.  It is then easy
to show using (\ref{comu1}), (\ref{comu2}) and (\ref{mono1}) that
\begin{gather*}
 \big [t(\lambda), E^{(N)}_{j} \big ] = \big [t(\lambda),
F^{(N)}_{j} \big ] =\big [t(\lambda), H^{(N)}_{j} \big ] =0,
\end{gather*} which proves that the transfer matrix for $K^{(L,
R)}={\mathbb I}$ is indeed ${\cal U}_{q}(sl_2) \otimes {\cal
U}_{{\mathrm i}}(sl_2)$ symmetric.

$\bullet$ {\it The ${\cal U}_{r}(sl_2)$, ${\cal U}_{\hat r}(sl_2)$
symmetries:} As discussed in \cite{our2} in addition to the ${\cal
U}_{q}(sl_2)$, ${\cal U}_{{\mathrm i}}(sl_2)$ symmetry the open twin
Hamiltonian also enjoys the ${\cal U}_{r}(sl_2) \otimes {\cal
U}_{\hat r}(sl_2)$ symmetry. We shall show that the full transfer
matric enjoys these symmetries as well. In particular, the following
actions of ${\cal U}_{\hat r}(sl_2)$, ${\cal U}_{r}(sl_2)$ on
${\mathbb C}^{4}$ were introduced, i.e. 
\begin{gather*}
 \rho_{1}({\cal H}) =
{\mathbb I} \otimes \hat r^{{1\over 2}\sigma^{z}}, \qquad \rho_{1}({\cal
E})={\mathbb I} \otimes  \sigma^{+},
\qquad \rho_{1}({\cal F})={\mathbb I} \otimes \sigma^{-}, \nonumber\\
\rho_{2}({\cal H})= r^{-{1\over 2}\sigma^{z}}\otimes {\mathbb I},
\qquad \rho_{2}({\cal F}) = \sigma^{+}\otimes {\mathbb I},
\qquad \rho_{2}({\cal E}) = \sigma^{-}\otimes {\mathbb I} . 
\end{gather*}
 Setting \begin{gather*}
  \tilde E_{j}^{(N)}= \rho_{j}^{\otimes N}
(\Delta^{(N)}({\cal E})),\qquad \tilde F^{(N)}_{j}=\rho_{j}^{\otimes N}
(\Delta^{(N)}({\cal F})) ,\\
\tilde H_{j}^{(N)}= \rho_{j}^{\otimes
N} (\Delta^{(N)}({\cal H})), \qquad  j\in \{1,  2 \} 
\end{gather*}
 it was
shown in \cite{our2} that 
\begin{gather*}
 \big [\tilde E_{j}^{(N)},
\Theta({\mathbb U}_{l}) \big ] = \big [\tilde F^{(N)}_{j},
\Theta({\mathbb U}_{l}) \big ] =  \big [\tilde H^{(N)}_{j},
\Theta({\mathbb U}_{l}) \big ] =0,\qquad l\in \{1, \ldots,
N-1\}.
\end{gather*}
 As in the ${\cal U}_{q}(sl_2)$, ${\cal
U}_{{\mathrm i}}(sl_2)$ cases generalized intertwining relations
between ${\cal T}$ (\ref{transfer1}) and the representations
$\rho_{j}$ introduced above are valid. Exploiting such relations we
may immediately obtain exchange relations given in Appendix B
(\ref{tcomu1}), (\ref{tcomu2}). In a straightforward manner, with
the use of (\ref{tcomu1}), (\ref{tcomu2}) and (\ref{mono1}) it may
be shown that {\samepage
\begin{gather*}
 \big [t(\lambda), \tilde E^{(N)}_{j} \big ] =
\big [t(\lambda), \tilde F^{(N)}_{j} \big ] =\big [t(\lambda),
\tilde H^{(N)}_{j} \big ] =0 
\end{gather*}
 which proves that the
transfer matrix is also ${\cal U}_{r}(sl_2) \otimes {\cal U}_{\hat
r}(sl_2)$ symmetric.}

It is f\/inally worth noting that the maps $\sigma_i$ can be in fact
expressed in terms of the representations $\rho_i$ in the following
manner 
\begin{gather*}
 \sigma_1({\cal E}) = \rho_1({\cal E}) \rho_2({\cal
F}), \qquad \sigma_1({\cal F}) = \rho_1({\cal F}) \rho_2({\cal E}),
\qquad \sigma_1({\cal H}) \sigma_2({\cal H})= (-1)^{{1\over 2}}
\rho_1({\cal H}) \rho_2({\cal H}),\\
 \sigma_2({\cal E}) =
\rho_1({\cal E}) \rho_2({\cal E}), \qquad \sigma_2({\cal F}) =
\rho_1({\cal F}) \rho_2({\cal F}). 
\end{gather*}

We have been able to show, for the moment, that the asymmetric twin
open transfer matrix (\ref{transfer1}) with trivial boundary
conditions is ${\cal U}_{q}(sl_2) \otimes {\cal U}_{\mathrm
i}(sl_2)$ (${\cal U}_{r}(sl_2) \otimes {\cal U}_{\hat r}(sl_2)$)
symmetric. Whether there exist further recognized symmetries
associated to the twin spin chain with trivial boundaries is still a
question under investigation. It should be stressed however that up
to date we have not been able to identify further charges,
associated to some familiar quantum algebra, commuting with the open
asymmetric twin transfer matrix (\ref{transfer1}).

\subsection[Non-trivial boundary conditions, $K^{(R)} \neq {\mathbb I}$]{Non-trivial boundary 
conditions, $\boldsymbol{K^{(R)} \neq {\mathbb I}}$}

It is also desirable to investigate the transfer matrix symmetry
when a non-trivial right bounda\-ry~(\ref{K3}), emerging from the
representations $\Theta^I$ (\ref{cabl})--(\ref{cabl3}), is
implemented. The left boundary is kept trivial i.e.\ $K^{(L)}=
{\mathbb I}$. Inspired basically by the symmetry of the open XXZ
spin chain with non-diagonal right boundary \cite{mene, dema, doikoun} 
we consider the combination of generators of the quantum
algebra ${\cal U}_{{\mathrm q}}({sl_2})$, ${\mathrm q}\in \{ q,
{\mathrm i}, r, \hat r\}$ 
\begin{gather*}
 {\cal Q}_{\mathrm q} = {\mathrm
q}^{-{1\over 2}} {\cal H}\ {\cal E}+{\mathrm q}^{ {1\over 2}}
{\cal H}\ {\cal F} +x_{{\mathrm q}} {\cal H}^{2}-x_{{\mathrm q}}
{\mathbb I} 
\end{gather*} the constants $x_{{\mathrm q}}$ will be
identif\/ied later on in this section. The charge ${\cal Q}_{\mathrm
q}$ is equipped with a~co-product structure, i.e., 
\begin{gather*}
 \Delta({\cal
Q}_{\mathrm q} ) = {\mathbb I} \otimes {\cal Q}_{\mathrm q} +{\cal
Q}_{\mathrm q}  \otimes {\cal H}^{2}. 
\end{gather*}
 As in the case
of trivial boundary conditions we shall state the following general
argument. Consider a solution of the Yang--Baxter equation
satisfying (\ref{ir}), and a $K^{(R)}$ matrix, solution of the
ref\/lection equation, satisfying 
\begin{gather}
 {\mathrm h}({\cal Q}_{\mathrm
q})\ K^{(R)}(\lambda)= K^{(R)}(\lambda)\ {\mathrm h}({\cal
Q}_{\mathrm q}). \label{irk}
\end{gather}
 Then using (\ref{ir}) and
(\ref{irk}) one can show in a straightforward fashion that
generalized intertwining relations hold also for the corresponding
${\cal T}$ (\ref{transfer1}) (see also \cite{doikoun, doikou2} for a
detailed proof) 
\begin{gather}
 ({\mathrm h} \otimes {\mathrm h}^{\otimes N})
\Delta^{'(N+1)}({\cal Q}_{\mathrm q})\ {\cal T}(\lambda) = {\cal
T}(\lambda)\ ({\mathrm h} \otimes {\mathrm h}^{\otimes
N})\Delta^{'(N+1)}({\cal Q}_{\mathrm q}). \label{itt2} 
\end{gather}
Again the
latter relations are of the type (\ref{it0}) for $\lambda' \to \pm
\infty$

We could have considered the combination of the quantum algebra
${\cal U}_{{\mathrm q}}(\widehat{sl_2})$ generators \cite{doikoun,
our2} and then exploit intertwining relations between the
co-products of the ${\cal U}_{{\mathrm q}}(\widehat{sl_2})$
generators and the operator ${\cal T}$, but as already discussed the
twin $R$ matrix does not satisfy any obvious intertwining relations
with the elements of ${\cal U}_{{\mathrm q}}(\widehat{sl_2})$,
therefore we focus on the non af\/f\/ine case. In what follows we shall
treat separately each one of the boundaries associated to
$\Theta^I$.

{\bf Type (i)} It was shown in \cite{our2} for the solution type (i)
(\ref{K3}), that
\begin{gather}
\sigma_{1} (x) K^{(R)}(\lambda)= K^{(R)}(\lambda) \sigma_{1} (x), \qquad
x \in {\cal U}_{{\mathrm i}}(sl_{2}),\nonumber\\
\sigma_{2} ({\cal Q}_q)\ K^{(R)}(\lambda)= K^{(R)}(\lambda)\
\sigma_{2} ({\cal Q}_q),  \label{cr1}
\end{gather}
 provided that 
 \begin{gather}
  x_{q} =
{Q -Q^{-1} \over q - q^{-1}}, \label{xi1} 
\end{gather}
  $\sigma_{j}$ are given
by (\ref{gen1}). The f\/irst of the equations (\ref{cr1}) implies that
the presence of the non-trivial boundary (i) does not break the
${\cal U}_{{\mathrm i}}(sl_2)$ symmetry, namely 
\begin{gather}
 \big [
t(\lambda), E^{(N)}_{1} \big ] =\big [ t(\lambda), F^{(N)}_{1}
\big ]=\big [ t(\lambda), K^{(N)}_{1} \big ]=0. \label{s0} 
\end{gather}
Using the second equation in (\ref{cr1}) it is clear that
intertwining relations (\ref{itt2}) hold for ${\mathrm h} \to
\sigma_{2}$. Setting also 
\begin{gather}
 {\mathbb Q}^{(N)}_q =
\sigma_{2}^{\otimes N}(\Delta^{(N)}({\cal Q}_q)) \label{set} 
\end{gather} we
obtain via (\ref{itt2}) for $\sigma_{2}$ and (\ref{mono1}) the
following exchange relations 
\begin{gather}
 \big [ {\mathbb Q}^{(N)}_q, {\cal
A} \big ] = \big [ {\mathbb Q}_q^{(N)}, {\cal D} \big ] =0, \qquad
\big[ {\mathbb Q}_q^{(N)}, {\cal A}_{1} \big ]= q({\cal B}_{5} -{\cal
C}_{5}), \nonumber\\
 \big [ {\mathbb Q}_q^{(N)}, {\cal A}_{2} \big
]=-q^{-1}({\cal B}_{5} -{\cal C}_{5}). \label{comu1b} 
\end{gather} Using the
latter relations and the form of the transfer matrix (\ref{mono1})
we can show that 
\begin{gather}
 \big [t(\lambda), {\mathbb Q}_q^{(N)} \big ]
=0. \label{s1} 
\end{gather} The boundary associated to solution (i) preserves
the ${\cal U}_{{\mathrm i}}(sl_2)$ symmetry (\ref{s0}), and also
preserves part of ${\cal U}_{q}(sl_{2})$, that is the charge
${\mathbb Q}_{q}^{(N)}$ (\ref{s1}).

{\bf Type (ii)} For the solution type (ii) (\ref{K3}) one has that
\cite{our2} 
\begin{gather}
\sigma_{2} (x) K^{(R)}(\lambda)=
K^{(R)}(\lambda)
\sigma_{2} (x), \qquad x \in {\cal U}_{q}(sl_2),\nonumber\\
\sigma_{1} ({\cal Q}_{{\mathrm i}}) K^{(R)}(\lambda)=
K^{(R)}(\lambda) \sigma_{1} ({\cal Q}_{{\mathrm i}}), \label{cr1b}
\end{gather}
 provided that 
 \begin{gather}
  x_{{\mathrm i}} = - {Q -Q^{-1} \over 2{\mathrm
i}}. \label{xi2}
\end{gather}
 The f\/irst of the equations (\ref{cr1b}) implies
that the presence of the non-trivial boundary (ii) does not break
the ${\cal U}_{q}(sl_2)$ symmetry, i.e. 
\begin{gather}
 \big [ t(\lambda),
E^{(N)}_{2} \big ] =\big [ t(\lambda), F^{(N)}_{2} \big ]=\big [
t(\lambda), H^{(N)}_{2} \big ]=0. \label{s0b} 
\end{gather}  It can be shown
using the second equation in (\ref{cr1b}) that generalized
intertwining relations (\ref{itt2}) are valid for ${\mathrm h} \to
\sigma_{1}$. Recalling (\ref{set}) and setting 
\begin{gather} {\mathbb
Q}^{(N)}_{{\mathrm i}} = \sigma_{1}^{\otimes N} (\Delta^{(N)}({\cal
Q}_{{\mathrm i}})) \label{ii} 
\end{gather} we obtain the following exchange
relations 
\begin{gather}
 \big [ {\mathbb Q}^{(N)}_{{\mathrm i}}, {\cal A}_{1}
\big ] = \big [ {\mathbb Q}^{(N)}_{{\mathrm i}}, {\cal A}_{2} \big
] =0, \qquad \big [ {\mathbb Q}^{(N)}_{{\mathrm i}}, {\cal A} \big ]=
{\mathrm i}^{-1}({\cal B} -{\cal C}), \nonumber\\
\big [ {\mathbb
Q}^{(N)}_{{\mathrm i}}, {\cal D} \big ]=-{\mathrm i}({\cal B}
-{\cal C}). \label{comu2b} 
\end{gather}
 From the latter relations and the
form of the transfer matrix (\ref{mono1}) we conclude that 
\begin{gather}
 \big
[t(\lambda), {\mathbb Q}^{(N)}_{{\mathrm i}} \big ] =0.
\label{s2}
\end{gather}
 The presence of boundary of type (ii) preserves the
${\cal U}_{q}(sl_2)$ symmetry (\ref{s0b}), and also the boundary
charge  ${\mathbb Q}^{(N)}_{{\mathrm i}}$ commutes with the transfer
matrix.

{\bf Type (+)} It is clear from the form of the solution (+)
(\ref{K3}), (\ref{cable+}) and taking into account relations
(\ref{itt2}) for ${\mathrm h} \to \sigma_{1}, \sigma_{2}$ (which
apparently also hold for $M^i$ (\ref{cabl}) and $M^{ii}$
(\ref{cabl2})), provided that relations (\ref{xi1}), (\ref{xi2})
hold simultaneously. Both sets of commutation relations
(\ref{comu1b}), (\ref{comu2b}) are  valid and therefore 
\begin{gather}
 \big
[t(\lambda), {\mathbb Q}^{(N)}_{{\mathrm i}} \big ] = \big
[t(\lambda), {\mathbb Q}^{(N)}_q \big ] =0. \label{symm0} 
\end{gather}
 The
presence of solution type (+) breaks both ${\cal U}_{q}(sl_2)$ and
${\cal U}_{\tilde q}(sl_2)$, and the remaining conserved quantities
are the boundary non-local charges ${\mathbb Q}^{(N)}_q$, ${\mathbb
Q}^{(N)}_{{\mathrm i}}$.

{\bf Type (iii)} Finally, for the solution type (iii) (\ref{K3}) the
following commutation relations are valid \cite{our2} 
\begin{gather*}
\rho_{1} ({\cal Q}_{\hat r}) K^{(R)}(\lambda)= K^{(R)}(\lambda)
\rho_{1} ({\cal Q}_{\hat r}),\nonumber\\
 \rho_{2} ({\cal Q}_{r})\
K^{(R)}(\lambda)= K^{(R)}(\lambda) \rho_{2} ({\cal Q}_{r})
\end{gather*}
 provided that the constants $x_{r}$, $x_{\hat r}$
are given by 
\begin{gather*}
 x_{r} = {\mathrm i}{\sqrt{{\mathrm i}Q} +
\sqrt{-{\mathrm i}Q^{-1}} \over  r -r^{-1}}, \qquad
x_{\hat r} =
{\sqrt{{\mathrm i}Q} - \sqrt{-{\mathrm i}Q^{-1}} \over \hat r -\hat
r^{-1}}. 
\end{gather*}
 Again we can show that (\ref{itt2}) are valid for both
$\rho_{1}$, $\rho_{2}$. Setting 
\begin{gather}
 {\mathbb Q}^{(N)}_{\hat r} =
\rho_{1}^{\otimes N}(\Delta^{N}({\cal Q}_{\hat r})), \qquad
{\mathbb
Q}^{(N)}_{r} = \rho_{2}^{\otimes N} (\Delta^{N}({\cal Q}_{r}))
\label{iii} 
\end{gather}
 and with the help of (\ref{itt2}) for ${\mathrm h}
\to \rho_{1}, \rho_{2}$ and (\ref{mono1}) we obtain exchange
relations of the type: 
\begin{gather*}
 r^{-1}\big [{\mathbb Q}_{r}^{(N)},
{\cal A} \big ] = -r\big [{\mathbb Q}_{r}^{(N)}, {\cal A}_2  \big ]
=({\cal C}_{2} -{\cal B}_{2}), \nonumber\\
r^{-1}\big [{\mathbb
Q}_{r}^{(N)}, {\cal A}_{1}  \big ] =-r\big [{\mathbb Q}_{r}^{(N)},
{\cal D} \big ] =({\cal C}_{3} -{\cal B}_{3}), \nonumber\\ 
  -\hat
r^{-1}\big [{\mathbb Q}_{\hat r}^{(N)}, {\cal A} \big ] = \hat r
\big [{\mathbb Q}_{\hat r}^{(N)}, {\cal A}_1 \big ] =({\cal C}_{1}
-{\cal B}_{1}), \nonumber\\
-\hat r\big [{\mathbb Q}_{\hat r}^{(N)}, {\cal
A}_{2}  \big ] =\hat r^{-1}\big [{\mathbb Q}_{\hat r}^{(N)}, {\cal
D} \big ] =({\cal C}_{4} -{\cal B}_{4}). 
\end{gather*}
 The latter
relations lead to the following, 
\begin{gather}
 \big [t(\lambda), {\mathbb
Q}^{(N)}_{\hat r} \big ] =\big [t(\lambda), {\mathbb Q}^{(N)}_{r}
\big ] =0. \label{symm} 
\end{gather}
  In this case both ${\cal U}_{q}(sl_2)$,
${\cal U}_{\tilde q}(sl_2)$ (${\cal U}_{r}(sl_2)$, ${\cal U}_{\hat
r}(sl_2)$) are broken, and the remaining conserved quantities are
the charges ${\mathbb Q}_{\hat r}^{(N)}$, ${\mathbb Q}_{r}^{(N)}$.

As in the case with trivial boundary conditions, discussed in the
previous section, the crucial point raised is whether there exist
further well known symmetries associated to the model under
consideration. Up to date we have not succeed to identify further
commuting quantities, associated to some familiar quantum algebra. A
pertinent question is whether the highly involved boundary non-local
charges ${\cal T}^{\pm}_{ab}$ introduced in Section~4 may be written
in terms of the more familiar charges (\ref{set}), (\ref{ii}),
(\ref{iii}). Presumably some of the non-local charges ${\cal
T}^{\pm}_{ab}$ may be expressed in terms of those but not all of
them. It is worth emphasizing that all the conserved quantities
found for both trivial and non trivial right boundary satisfy
algebraic relations of the type (\ref{it0}), with no spectral
dependance. Finally in the case of a non trivial left boundary one
has to exploit exchange relations involving all the entries of the
${\cal T}$ matrix and then extract the appropriate combination of
non-local charges commuting with transfer matrix.

\section{Discussion}

Let us now brief\/ly review the main f\/indings of the present article.
The main objective of this work was the study of the symmetry of the
open asymmetric twin chain. In this spirit we were able to show that
the boundary quantum algebra provides a symmetry for any open
transfer matrix with trivial boundary (\ref{symme2}). It was also
shown that the boundary quantum algebra is in addition the
centralizer of the $B$ type Hecke algebra (\ref{a}). Furthermore we
derived sets of convenient exchange relations for the asymmetric
twin model with both trivial and non trivial boundaries emerging
from the generalized intertwining relations (\ref{itt}),
(\ref{itt2}). By exploiting such relations we proved the commutation
of the transfer matrix with certain non-local charges. More
precisely, in the case of trivial boundaries the derived conserved
charges consist tensor representations of ${\cal U}_q(sl_{2})\otimes
{\cal U}_{{\mathrm i}}(sl_{2})$ (${\cal U}_r(sl_{2})\otimes {\cal
U}_{\hat r}(sl_{2})$). When a non-trivial right boundary is
implemented the symmetry of the transfer matrix as expected is
reduced (\ref{s0}), (\ref{s1}), (\ref{s0b}), (\ref{s2}),
(\ref{symm0}), (\ref{symm}). Depending on the choice of boundaries
some of the symmetry is preserved, and the new conserved quantities
are expressed as combinations of the generators of the
aforementioned quantum algebras. Notice that we essentially extend
the results of \cite{our2} in as much as the results of \cite{pasa}
are extended in \cite{kusk}. As already pointed out the discovered
`familiar' symmetries do not seem to consist the full symmetry of
the transfer matrix in \cite{our2}. The full symmetry (non-Abelian)
of the model is presumably the boundary quantum algebra, but at this
stage this is rather a conjecture, which needs to be further
checked.

The relation between the boundary non-local charges and the spectrum
and Bethe ansatz equations is also an intricate problem. Usually
such relations emerge from the asymptotic behavior of the ${\cal T}$
matrix, which as already mentioned is not at all straightforward for
the asymmetric twin model. More precisely, it is clear that the
asymptotics of the transfer matrix may be expressed as (see also
e.g.~\cite{done2, doikou3}) 
\begin{gather*} t(\lambda \to \pm \infty) \propto
\sum_{a=1}^{4}{\cal T}^{\pm}_{aa},
\end{gather*} hence the spectrum of ${\cal
T}_{aa}^{\pm}$ will provide consequential information regarding the
asymptotic behaviour of the spectrum of the asymmetric twin chain.
The key point is to derive the explicit form of the objects ${\cal
T}_{aa}^{\pm}$, and express them, if possible, in terms of the
familiar non-local charges of Sections~5.2, 5.3.  For the moment
there is no apparent link between the conserved quantities and the
spectrum and this is the main obstacle in deriving the spectrum of
the twin transfer matrix (see also \cite{our2}). However, in
\cite{our2} the equivalence of the spectrum of the open twin and XXZ
Hamiltonians was established, and consequently the form of the
spectrum of the asymmetric twin chain was derived. It is worth
pointing out that the diagonalization of the non-local charges is
also a particularly challenging problem, and it has been already
solved for the XXZ model for particular representations in
\cite{base,base1,nire,nire1}. In fact the boundary non-local charges of
Section~5.3 having exactly the same structure as the ones of the
XXZ model may be diagonalized along the lines described in
\cite{base,base1,nire,nire1}. Finally, an interesting point to pursue
is the generalization of the asymmetric twin representation
consisting of representations associated to higher rank algebras. We
hope to report on these matters in a forthcoming work.

\appendix
\section{Appendix}

We present here $\Theta({\mathbb U}_{1})$ as a $16 \times 16$ matrix
acting {\em not} on ${\mathbb V}_{2^-} \otimes {\mathbb V}_{1^-}
\otimes {\mathbb V}_{1^+} \otimes {\mathbb V}_{2^+}$, but on $(
{\mathbb V}_{1^-} \otimes {\mathbb V}_{1^+} ) \otimes ( {\mathbb
V}_{2^-} \otimes {\mathbb V}_{2^+} )= {\mathbb V}_{\tilde 1} \otimes
{\mathbb V}_{\tilde 2}$, according to the space/mirror space
notation (see also~(\ref{fold}))
\[
\Theta({\mathbb U}_{1})= \left( \begin{array}{cccccccccccccccc}
0&&&&&&&&&&&&&&& \\
&0&&&&&&&&&&&&&& \\
&&0&&&&&&&&&&&&& \\
&&&-i^{}&&&-r^{-1}&&&-\hat r^{}&&&1&&& \\
&&&&0&&&&&&&&&&& \\
&&&&&0&&&&&&&&&& \\
&&&-r^{-1}&&&-q^{-1}&&&1&&&-\hat r^{-1}&&& \\
&&&&&&&0&&&&&&&& \\
&&&&&&&&0&&&&&&& \\
&&&-\hat r^{}&&&1&&&-q&&&-r^{}&&& \\
&&&&&&&&&&0&&&&& \\
&&&&&&&&&&&0&&&& \\
&&&1&&&-\hat r^{-1}&&&-r^{}&&&i&&& \\
&&&&&&&&&&&&&0&& \\
&&&&&&&&&&&&&&0& \\
&&&&&&&&&&&&&&&0 \\
\end{array}\right)\otimes  \cdots \otimes {\mathbb I}.
\]

\section{Appendix}

In this Appendix exchange relations arising from the generalized
intertwining relations are reported. First we present exchange
relations involving the entries of ${\cal T}$ and representations of
the quantum algebra ${\cal U}_{q}(sl_{2})$. Let $[X, Y]_{{\mathrm
q}} = XY-{\mathrm q} YX$, then 
\begin{gather}
 \big [{\cal A}, H^{(N)} \big ]
=0, \qquad \big [{\cal D}, H^{(N)} \big ] =0, \nonumber\\
\big [{\cal C},
(H^{(N)})^{\pm 1} \big ]_{q^{\mp 1}}=0, \qquad
\big [{\cal B},(H^{(N)})^{\pm 1} \big ]_{q^{\pm 1}}=0, \label{co1}\\
 \big [E^{(N)}, {\cal A} \big ] = -q^{-{1\over 2}} (H^{(N)})^{-1} {\cal
C},\qquad
\big [ E^{(N)},{\cal D} \big ] = q^{{1\over 2}}  {\cal C} (H^{(N)})^{-1}, \nonumber\\
 \big [E^{(N)}, {\cal C} \big ]_{q} =0, \qquad \big [E, {\cal B}\big
]_{q^{-1}}= q^{-{1\over 2}} \big ({\cal A} (H^{(N)})^{-1}-
(H^{(N)})^{-1} {\cal D} \big ), \label{co2}\\
 \big [
F^{(N)}, {\cal A} \big ] = q^{-{1\over 2}} {\cal B}
(H^{(N)})^{-1},\qquad
\big [ F^{(N)}, {\cal D} \big ] = -q^{{1\over
2}} (H^{(N)})^{-1} {\cal B},\nonumber\\
  \big [F^{(N)}, {\cal B}
\big ]_{q^{-1}} =0, \qquad \big [F^{(N)}, {\cal C}\big ]_{q}= q^{{1\over
2}} \big ({\cal D} (H^{(N)})^{-1}-(H^{(N)})^{-1} {\cal A}\big
).\label{co3} 
\end{gather}
 Exchange relations between the entries of ${\cal
T}$ and representations of the quantum algebras ${\cal
U}_{q}(sl_{2})$, ${\cal U}_{{\mathrm i}}(sl_{2})$ are given below
\begin{gather}
 \big [ H^{(N)}_{j}, {\cal A} \big ]= \big [ H^{(N)}_{j},
{\cal D} \big
]= \big [ H_{j}^{(N)}, {\cal A}_{i} \big ]=0, \qquad i,j\in \{1, 2\}, \nonumber\\
\big [ {\cal C}, (H_{1}^{(N)})^{-1} \big ]_{\mathrm i}=\big [
{\cal B}, (H_{1}^{(N)})^{-1} \big ]_{{\mathrm i}^{-1}}=\big [ {\cal
C}_{5}, (H_{2}^{(N)})^{-1} \big ]_{ q^{-1}}=\big [ {\cal B}_{5},
(H_{2}^{(N)})^{-1} \big ]_{q}=0 ,\label{comu1} \\
 \big [
E^{(N)}_{1}, {\cal A} \big ] =-{\mathrm i}^{-{1\over
2}}(H^{(N)}_{1})^{-1} {\cal C}, \qquad \big [ E^{(N)}_{1}, {\cal D} \big
] ={\mathrm i}^{{1\over 2}}{\cal C}\ (H_{1}^{(N)})^{-1}, \qquad \big [
E^{(N)}_{1}, {\cal A}_{j} \big ]=0, \nonumber\\ 
\big [F^{(N)}_{1},
{\cal A} \big ] ={\mathrm i}^{-{1\over 2}}{\cal B}
(H_{1}^{(N)})^{-1} , \qquad \big [ F^{(N)}_{1}, {\cal D} \big ]
=-{\mathrm i}^{{1\over 2}} (H_{1}^{(N)})^{-1}{\cal B}, \qquad \big [
F^{(N)}_{1}, {\cal A}_{j} \big ]=0, \nonumber\\ 
\big [ F^{(N)}_{2},
{\cal A}_{1} \big ] =- q^{{1\over 2}}(H_{2}^{(N)})^{-1} {\cal
C}_{5}, \qquad \big [ F^{(N)}_{2}, {\cal A}_{2} \big ] =q^{-{1\over
2}}{\cal C}_{5} (H_{1}^{(N)})^{-1}, \nonumber\\
\big [ F^{(N)}_{2}, {\cal A}
\big ]=0, \qquad \big [ F^{(N)}_{2}, {\cal D} \big ]=0,\nonumber\\ 
\big [
E^{(N)}_{2}, {\cal A}_{1} \big ] =q^{{1\over 2}}{\cal B}_{5}
(H_{2}^{(N)})^{-1} , \qquad \big [ E_{2}^{(N)}, {\cal A}_{2} \big ]
=-q^{-{1\over 2}} (H_2^{(N)})^{-1}{\cal B}_{5}, \nonumber\\
\big [
E^{(N)}_{2}, {\cal A} \big ]=0, \qquad \big [ E^{(N)}_{2}, {\cal D} \big
]=0. \label{comu2} 
\end{gather}
 Finally exchange relations involving
the entries of ${\cal T}$ and representations of the quantum
algebras ${\cal U}_{r}(sl_{2})$, ${\cal U}_{\hat r}(sl_{2})$ are
presented below 
\begin{gather}
 \big [ {\cal B}_{1}, (\tilde
H_{1}^{(N)})^{-1} \big ]_{\hat r^{-1}}=\big [ {\cal B}_{4}, (\tilde
H_{1}^{(N)})^{-1} \big ]_{\hat r^{-1}}= \big [ {\cal C}_{1},
(\tilde H_{1}^{(N)})^{-1} \big ]_{\hat r} =\big [
{\cal C}_{4}, \tilde (H_{1}^{(N)})^{-1} \big ]_{\hat r} =0 ,\nonumber\\
\big [ {\cal B}_{2}, (\tilde H_{2}^{(N)})^{-1} \big ]_{r} =\big [
{\cal B}_{3}, (\tilde H_{2}^{(N)})^{-1} \big ]_{r}=\big [ {\cal
C}_{2}, (\tilde H_{2}^{(N)})^{-1} \big ]_{r^{-1}}= \big [ {\cal
C}_{3}, (\tilde H_{2}^{(N)})^{-1} \big ]_{r^{-1}} =0,
\label{tcomu1}\\
\big [\tilde E^{(N)}_{1}, {\cal A} \big ]=
-\hat r^{-{1\over 2}}(\tilde H_{1}^{(N)})^{-1}{\cal C}_{1},\qquad
\big
[\tilde E^{(N)}_{1}, {\cal A}_{1} \big ]= \hat r^{{1\over 2}}{\cal
C}_{1}(\tilde H_{1}^{(N)})^{-1}, \nonumber\\
\big [\tilde E^{(N)}_{1}, {\cal
A}_{2} \big ]= -\hat r^{-{1\over 2}}(\tilde H_{1}^{(N)})^{-1}{\cal
C}_{4}, \qquad
 \big [\tilde E^{(N)}_{1}, {\cal D} \big ]= \hat
r^{{1\over 2}}{\cal C}_{4}(\tilde H_{1}^{(N)})^{-1},\nonumber\\
\big [\tilde
F^{(N)}_{1}, {\cal A} \big ]= \hat r^{-{1\over 2}}{\cal B}_{1}
(\tilde H_{1}^{(N)})^{-1},\qquad
\big [\tilde F^{(N)}_{1}, {\cal A}_{1}
\big ]= -\hat r^{{1\over 2}} (\tilde H_{1}^{(N)})^{-1}{\cal B}_{1},
\nonumber\\ 
 \big [\tilde F^{(N)}_{1}, {\cal A}_{2} \big ]= \hat
r^{-{1\over 2}}{\cal B}_{4}(\tilde H_{1}^{(N)})^{-1}, \qquad
\big
[\tilde F^{(N)}_{1}, {\cal D} \big ]= r^{{1\over 2}}{\cal B}_{2}
(\tilde H_{2}^{(N)})^{-1},\nonumber\\
\big [\tilde E^{(N)}_{2}, {\cal A}
\big ]= r^{{1\over 2}}{\cal B}_{2}(\tilde H_{1}^{(N)})^{-1}, \qquad
 \big [\tilde E^{(N)}_{2},\ {\cal A}_{1} \big ]= r^{{1\over
2}}{\cal B}_{3} (\tilde H_{1}^{(N)})^{-1}, \nonumber\\
 \big [\tilde
E_{2}^{(N)}, {\cal A}_{2} \big ]= - r^{-{1\over 2}}(\tilde
H_{2}^{(N)})^{-1}{\cal B}_{2}, \qquad
\big [\tilde E^{(N)}_{2}, {\cal
D} \big ]= - r^{-{1\over 2}}(\tilde H_{1}^{(N)})^{-1}{\cal B}_{3}
\nonumber\\ 
 \big [\tilde F_{2}^{(N)}, {\cal A} \big ]= - r^{{1\over
2}}(\tilde H_{2}^{(N)}){\cal C}_{2}, \qquad
\big [\tilde F^{(N)}_{2},
{\cal A}_{1} \big ] = -r^{{1\over 2}}(\tilde H_{2}^{(N)})^{-1}{\cal
C}_{3},\nonumber\\
\big [\tilde F^{(N)}_{2}, {\cal A}_{2} \big ]=
r^{-{1\over 2}}{\cal C}_{2}(\tilde H_{2}^{(N)})^{-1}, \qquad 
 \big
[\tilde E^{(N)}_{2}, {\cal D} \big ]= r^{-{1\over 2}}{\cal C}_{3}
(\tilde H_{2}^{(N)})^{-1}. \label{tcomu2} 
\end{gather}

\subsection*{Acknowledgments} 
I am thankful to P.P.~Martin for
useful discussions. This work is supported by INFN, and the European
Network `EUCLID'; `Integrable models and applications: from strings
to condensed matter', contract number HPRN--CT--2002--00325.

\pdfbookmark[1]{References}{ref}
\LastPageEnding

\end{document}